\documentclass[fleqn,usenatbib]{mnras}

\usepackage{newtxtext,newtxmath}

\usepackage[T1]{fontenc}

\DeclareRobustCommand{\VAN}[3]{#2}
\let\VANthebibliography\thebibliography
\def\thebibliography{\DeclareRobustCommand{\VAN}[3]{##3}\VANthebibliography}

\usepackage{graphicx}	
\usepackage{amsmath}	

\title[Recurrent QPPs in GJ 3512]{A recurrent 70--100 minute quasi-periodic pulsation in the intermediate-aged mid-M dwarf GJ 3512}

\author[J. Lopez-Santiago et al.]{
J. L\'opez-Santiago,$^{1}$\thanks{E-mail: jalopezs@ing.uc3m.es}
F. Reale,$^{2,3}$
G. Micela,$^{2}$
L. Martino,$^{4}$
G. V\'azquez-Vilar$^{1}$
and J. Miguez$^{1}$
\\
$^{1}$Universidad Carlos III de Madrid, Department of Signal Theory and Communications, Avenida de la Universidad 30, 28911, Legan\'es (Madrid), Spain\\
$^{2}$INAF-Osservatorio Astronomico di Palermo, Piazza del Parlamento 1, 90134 Palermo, Italy\\
$^{3}$Dipartimento di Fisica \& Chimica, Universit\'a di Palermo, Piazza del Parlamento 1, 90134 Palermo, Italy\\
$^{4}$Universit\`a degli Studi di Catania, Corso Italia 55, Catania, Italy
}

\date{Accepted XXX. Received YYY; in original form ZZZ}

\pubyear{2025}

\begin{document}
\label{firstpage}
\pagerange{\pageref{firstpage}--\pageref{lastpage}}
\maketitle

\begin{abstract}
We report the discovery of a {recurrent} quasi-periodic pulsation (QPP) in the late-M dwarf GJ 3512 (M5.5V) using multiple \emph{TESS} datasets. A strong signal with a period of 70--100 minutes was detected in wavelet analyses of the two-minute cadence light curve from Sector 20. This signal was detected also in observations from Sectors 47 and 60. The QPP persisted for weeks in sector 20 and spanned nearly three years of \emph{TESS} coverage. There was no significant damping between major flares. This behavior contrasts with that of previously reported stellar QPPs, which are confined to individual flares and decay on timescales of minutes to hours.
The oscillation amplitude is at the milli-magnitude level. A pulsation origin is discarded since theoretical instability strips for 100-minute pulsations are restricted to pre-main sequence stars, while GJ 3512 is an intermediate age (2-8 Gyr) main-sequence dwarf. The persistence across independent \emph{TESS} sectors discards an instrumental artifact origin and points to a likely coronal origin instead, such as oscillatory reconnection or thermal non-equilibrium cycles in large active regions.
%
%
This represents the first detection of a  {likely} sustained QPP with these characteristics in a late-type star, highlighting the need for further investigation into physical mechanisms behind such variability.
\end{abstract}

\begin{keywords}
stars: late-type -- stars: magnetic fields -- stars: activity -- stars: coronae -- stars: flare -- stars: oscillations
\end{keywords}



\section{Introduction}

Quasi-periodic pulsations (QPPs) are a common manifestation of magnetohydrodynamic (MHD) processes in solar and stellar flares. In the Sun, QPPs have been detected across the electromagnetic spectrum, with periods ranging from sub-seconds to tens of minutes, and are typically interpreted as oscillatory signatures of MHD wave modes in coronal loops or as the result of repetitive or oscillatory magnetic reconnection \citep[e.g.,][]{Nakariakov2009,Kupriyanova2020}. Quasi-periodic pulsations have been increasingly reported with \emph{Kepler} and \emph{TESS}, particularly in active M dwarfs, where flare amplitudes are larger and more frequent than in other solar-type stars. Nevertheless, the vast majority of stellar QPP detections to date are confined to individual flares, lasting only for a few oscillation cycles (minutes to hours), with power that decays rapidly after the flare peak.

Stellar QPPs detected with \emph{Kepler} span periods between approximately 5 and 90 minutes \citep[e.g.,][]{Pugh2015,Balona2015}, while \emph{TESS} studies have extended this range to $\sim$10--70 minutes in M3--M4 dwarfs \citep{Ramsay2021}. In the only reported late-M case, Proxima Centauri (M5.5V) has shown hour-scale QPPs during two individual flares \citep{Vida2019}. Outside of these flare-associated detections, no long-lived QPPs have yet been reported in main-sequence M dwarfs. The phenomenon is therefore considered to be a transient, flare-localized signature.

Theoretical studies have explored the possibility that low-mass stars and brown dwarfs might pulsate globally. Non-adiabatic stability analyses \citep{RodriguezLopez2019} have predicted that late-M dwarfs could, in principle, support pulsations with amplitudes of milli-magnitudes and periods of 20--200 minutes. However, the corresponding instability strips are restricted to the pre-main sequence (PMS), associated to deuterium burning or convective blocking, and do not overlap with the main sequence. For older, fully convective late-M dwarfs, no viable pulsation-driving mechanism is expected. Thus, while the periods predicted for PMS pulsations overlap with the typical timescales of flare QPPs, the physical origin  {is entirely distinct}.

In this context, we present the discovery of a {recurrent, long-lived}, QPP in the late-M dwarf GJ~3512 (M5.5V) observed with \emph{TESS}. 
 {GJ~3512 is an intermediate-aged (2--8 Gyr) main-sequence star hosting an exoplanetary system \citep{Morales2019,Morales2025revisiting}. The two confirmed planets have orbital periods of approximately $200$ and $1600$~d and the star shows a rotation period of 84~d from photometric data \citep{LopezSantiago2020}. }
A strong oscillatory signal with a period between 70 and 100 minutes is detected in wavelet analyses, persisting across 20 days of monitoring. The oscillation exhibits moderate period drift and low amplitude variations, but remains coherent throughout the entire observation. Unlike previously reported stellar QPPs, which are confined to single flares, the oscillation in GJ~3512 is sustained on week-long timescales. Flare peaks are phase-coincident with QPP maxima, suggesting a causal connection between flaring activity and the global oscillation, yet the signal does not decay between flares. Similar behavior has been observed during another two \emph{TESS} observations from Sectors~47 and 60. Nevertheless, in Sector~60, the QPP signal is  {not confidently detected, possibly because of a decrease in the activity of the star, although an intense white light flare is detected at the end of the observations}.

We suggest the QPP is triggered in a long-lived active region hosting extended magnetic loop systems, which likely undergo continuous microflaring and reconnection-driven oscillations. Candidate mechanisms include oscillatory reconnection in a persistent arcade \citep{Kumar2025,Schiavo2024} and thermal non-equilibrium cycles in large-scale coronal structures \citep[e.g.][]{Froment2020}. The interpretation in terms of global pulsations, like p- or g-modes, is discarded. GJ~3512 is  {a main-sequence star} located far outside the PMS instability strips. Star-planet interaction {(SPI)} origin is also discarded because the star's closest companion, GJ~3512b, is not in close proximity to the star.  {Theoretical models show that magnetic interaction strength decreases steeply with orbital distance and becomes negligible beyond a few stellar radii \citep{cuntz2000stellar,ip2004star,zarka2007plasma}. Observational studies likewise detect SPI signatures only in very close-in hot Jupiters \citep[e.g.][]{shkolnik2005hot,2014pillitteriHD189733}}
The detection of the QPP in observations separated by approximately two years suggests that the signal does not originate from a single active region, as this would require an unusually long-lived feature that coincidentally remained visible during each observation.  {Long-lived active regions are expected on fully convective M dwarfs, where spot complexes and active longitudes can persist for tens to hundreds of days \citep[e.g.][]{giles2017kepler}.}

To our knowledge, this is the first reported detection of a sustained QPP in a late-M star, distinct from the flare-localized pulsations observed in previous \emph{Kepler} and \emph{TESS} studies. This discovery provides evidence that late-M dwarfs can support long-lived coronal oscillations, with implications for magnetic energy release, microflaring activity, and coronal heating in the most numerous stellar population in the Galaxy. 

\section{Observations and data analysis}

\begin{figure}
   \centering
   \includegraphics[width=\columnwidth]{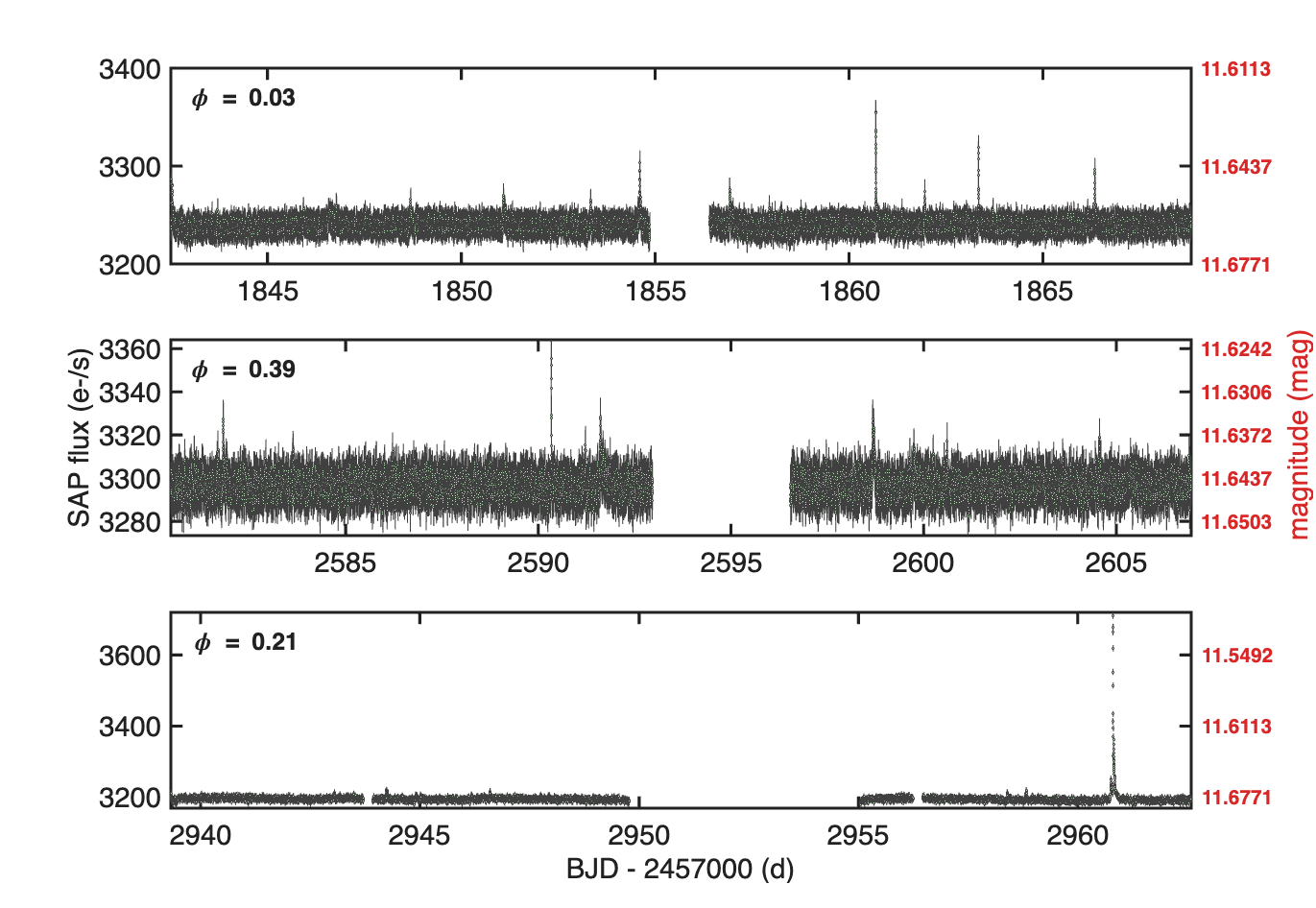} 
   \caption{\emph{TESS} PDCSAP light curves from Sectors~20, 47, and 60 for the star GJ~3512.  {The right axis refers to magnitudes in the \emph{TESS}  band (https://tess.mit.edu/public/tesstransients/pages/readme.html\#flux-calibration), which is close to the Cousin $I_C$ filter}. {The orbital phase of GJ~3512b (assuming $\phi = 0$ at the time of periastron passage) is given in each panel.}}
   \label{fig:TESSdatasets}
\end{figure}

We analyzed photometric data obtained in the 2-min cadence mode from the Transiting Exoplanet Survey Satellite (\emph{TESS}; \citealt{Ricker2015}). GJ~3512 was observed in three separate sectors: Sector~20 (BJD 2458842.5--2458862.5, 20 days), Sector~47 (BJD 2459213.1--2459233.1, 20 days), and Sector~60 (BJD 2459585.0--2459605.0, 20 days). All the data used in this paper can be found in MAST (http://dx.doi.org/10.17909/gsea-a830). These sectors provided nearly three years of time coverage between 2019 and 2022. For each sector, we retrieved the simple aperture photometry (SAP) and pre-search data conditioned simple aperture photometry (PDCSAP) light curves from the Mikulski Archive for Space Telescopes (MAST). We adopted the PDCSAP fluxes, which are corrected for common instrumental systematics. Outliers were identified as points deviating by more than $5\sigma$ from a running median (30-minute window) and were removed. Figure~\ref{fig:TESSdatasets} shows the three datasets. \emph{TESS} light curves contain short gaps due to scheduled data downlinks at perigee, which occur approximately every 13.7 days, and momentum dump events. During these events, science operations are briefly interrupted. These gaps are instrumental in origin and are not related to the astrophysical signal.

\begin{figure} 
   \centering
   \includegraphics[width=\columnwidth]{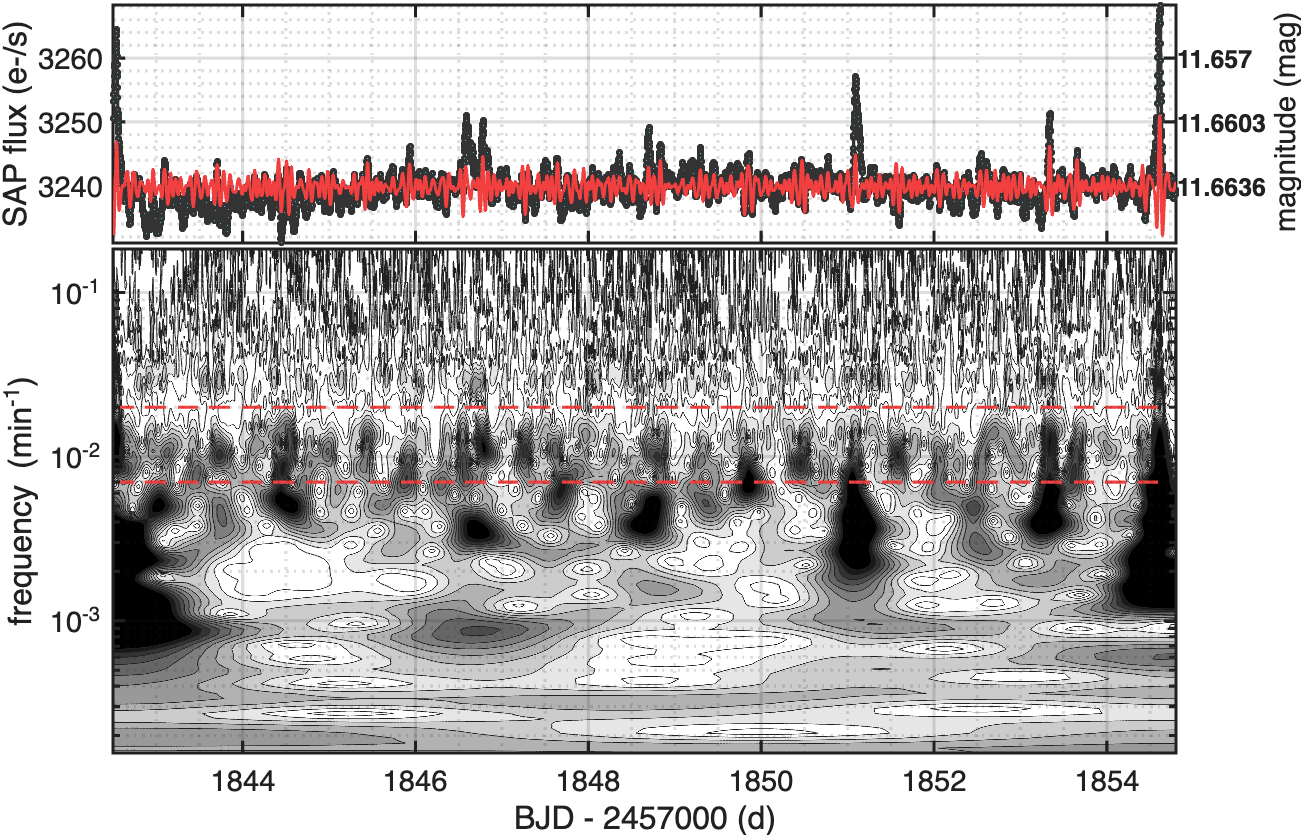} 
   \includegraphics[width=\columnwidth]{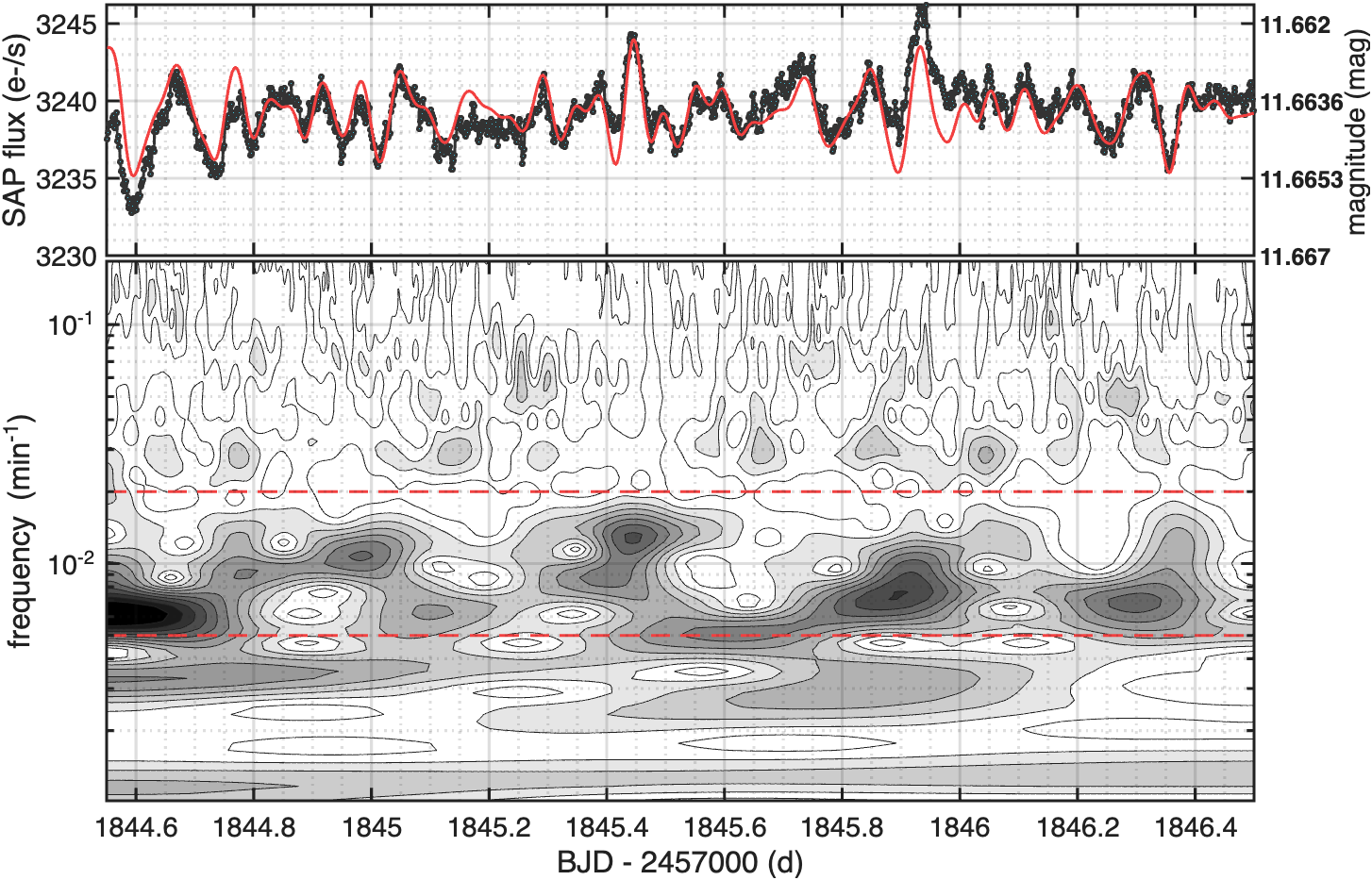}
   \caption{\textbf{Top}: Scalogram for the first $\sim13$ days of the TESS observation from December 24, 2019, to January 29, 2020. The dashed red lines in the scalogram indicate the frequency range selected for filtering. The red continuous line in the top panel is the filtered light curve. \textbf{Bottom}: Zoom of the figure in the left panel. Two days of observations are shown. Here, the QPP is clearly visible.}
   \label{fig:wavelet}
\end{figure}

To search for quasi-periodic signals, we first smoothed the extracted light curves using a 15-point moving mean ($\sim$30 minutes) to eliminate high-frequency noise.  {Note that the moving mean is equivalent to a first-order low-pass filter and cannot introduce periodic features into the time series \citep[see][for a formal description]{oppenheim1997signals}.} Then, we performed a continuous wavelet transform using the Morlet mother wavelet \citep{Torrence1998}. For simplicity and clarity, we present only the scalogram derived from Sector~20 here (Fig.~\ref{fig:wavelet}), as it provides a clear example. In this dataset, the oscillatory pattern remains present throughout the approximately 13 days shown in the figure, with slight variability in the period within the 70–100 minute range. Consistent with the response of a moving-average low-pass filter, the QPP becomes visible for modest windows ($N \ge 5$) as high-frequency noise is suppressed, but is attenuated for larger windows ($N \sim 30-40$), where the filter increasingly suppresses power at periods comparable to the window length. 
Similar behavior is observed in the other datasets. Figure~\ref{fig:wavelet} shows the scalogram resulting from the wavelet analysis (bottom panels) and the light curve during the observation period prior to the gap caused by the mission's perigee passage (top panels). The two dashed lines on the bottom panel indicate the frequency range in which the primary QPP feature is observed. Features in the scalogram that extend from high to low frequencies are caused by intense flares. We did not remove flare events prior to the analysis because the oscillatory signal appears both during and between flares. Nevertheless, their signatures in the scalogram are distinguishable as steep ridges extending from high to low frequencies.

We filtered out the other frequencies and reconstructed the light curve. The reconstruction was obtained by inverse-transforming the wavelet transform within the 70–100 minute band, after applying the cone-of-influence mask to exclude edge effects. The result is shown in red in the top panel of the figure. 

\section{Statistical significance of the signal}

\begin{figure}
   \centering
   \includegraphics[width=\columnwidth]{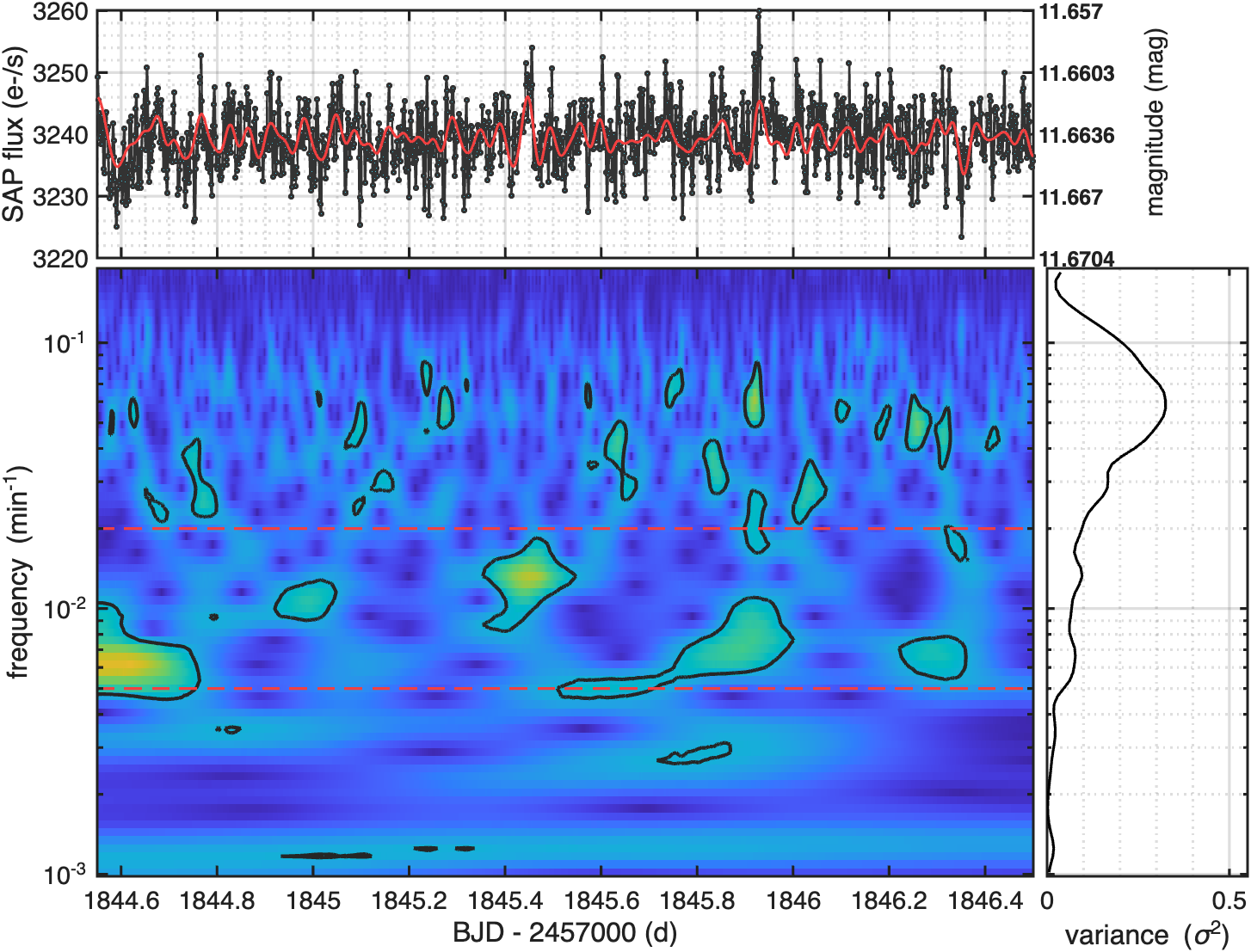} 
   \caption{Wavelet scalogram ($|S|$) per 99.9\% confidence threshold ($|S|$/threshold), for GJ~3512. The dashed red lines in the scalogram indicate the frequency range selected for filtering, which correspond to periods 70--120 minutes. The black continuous lines correspond to the 99.9\% confidence threshold.}
   \label{fig:GJ3512levels}
\end{figure}

 {The significance of the QPP was assessed in the wavelet scalogram using surrogate-data tests. Specifically, we generated surrogate light curves by randomly permuting the observed samples, which preserves the marginal distribution while suppressing temporal correlations \citep{theiler1992testing}. This ensemble defines the null hypothesis and was used to derive confidence levels in the scalogram \citep{schreiber2000surrogate}. A total of 10000 surrogate curves were generated and confidence thresholds were derived from them. Figure~\ref{fig:GJ3512levels} shows the result of dividing the scalogram (from the non-smoothed light-curve) in Figure~\ref{fig:wavelet} (bottom panel) by the 99.9\% confidence threshold. This confidence threshold is also shown as continuous black curves. Note that, with this methodology, the thresholds are derived for each pair time-frequency in the scalogram. Features with intensities larger than that threshold are significant at $<0.1$\% level ($p \leq 0.001$).
}

\begin{figure}
   \centering
   \includegraphics[width=\columnwidth]{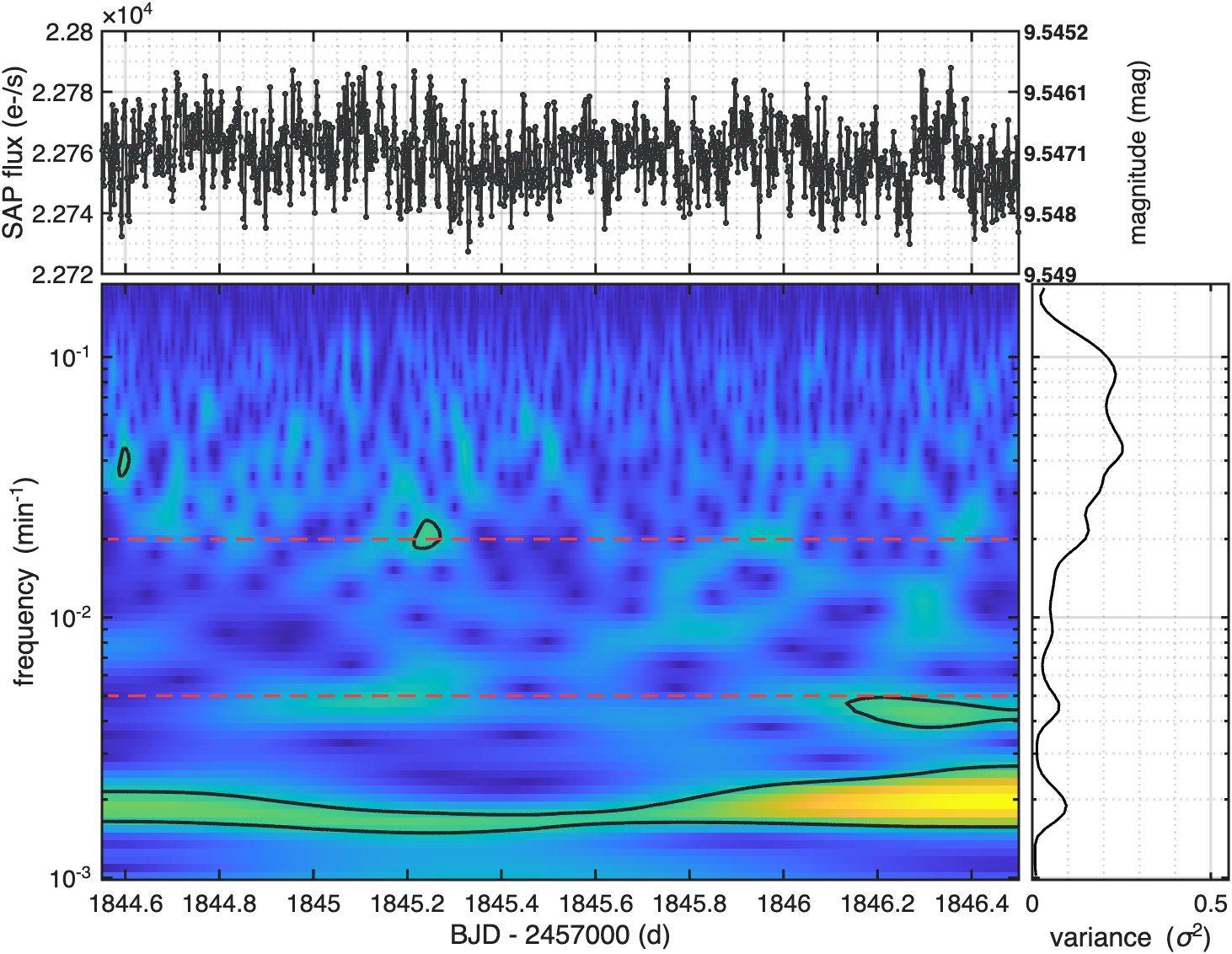} 
   \caption{Wavelet scalogram ($|S|$) per 99.9\% significance threshold ($|S|$/threshold), for TYC~3804-1138-1. The dashed red lines in the scalogram indicate the frequency range selected for filtering, which correspond to periods 70--120 minutes. The black continuous lines correspond to the 99.9\% significance threshold.}
   \label{fig:TYClevels}
\end{figure}

 {To discard a possible instrumental artifact, we selected a bright source in the same CCD and chip of the observations in Sector~20 and reproduced the wavelet analysis performed with GJ~3512. There are several bright sources detected in the same chip. We chose TYC~3804-1138-1, a moderately {bright} star \citep[V=10.1 mag;][]{Hog2000Tycho} showing no stellar activity during that \emph{TESS} observation. We first extracted its light curve following the same procedure than for GJ~3512 and derived significance thresholds as previously explained. The results are shown in Fig.~\ref{fig:TYClevels} for the same time period than in Fig~\ref{fig:GJ3512levels}, where the color-coded image corresponds to the scalogram of the signal divided by the 99.9\% significance threshold. No significant signal is detected in the frequency range corresponding to the period 70--120 min. Only a low-frequency signal is detected in this observing time period. This result is consistent with the 70--120~min QPP observed in GJ~3512 being an intrinsic signal from the source. 
} 

\section{Results}

The wavelet analysis of the \emph{TESS} light curves reveals clear quasi-periodic modulation with a period ranging from approximately 70 to 100 minutes. In Sector~20 (see Fig.~\ref{fig:wavelet}), the oscillatory pattern remains visible throughout the 20-day period (with brief interruptions due to instrumental gaps). The reconstructed light curve, obtained by filtering the wavelet coefficients within this frequency band, shows a modulation with a peak-to-peak amplitude of approximately 1-2 mmag superimposed on the underlying stellar flux. A similar signal is present in Sector~47 and,  {marginally}, in Sector~60, which are separated by approximately one and two years, respectively, from the initial observation.  {Note that the activity of GJ~3512 {likely} decreases at the epoch of the observations in Sector~60, {as suggested by} the detection of very few flaring events (barely five events).} This persistence suggests that the phenomenon is not associated with a single transient event, but rather reflects a long-lived or recurrent process on the star. The oscillation exhibits slight variability in period in all three datasets but it does not show evidence of systematic damping over the period of observation. 

Flare activity is frequent, with an average of approximately one flare per day and one intense flare every two days. This is consistent with previous studies of mid-to-late M dwarfs \citep[see,][for recent results]{Davenport2019,Gunther2020,Stelzer2022}. Interestingly, there appears to be a correlation between the flare peaks and the maxima of the reconstructed QPP signal, although this result should be interpreted with caution. No other coherent signals were detected at periods longer than two hours, although a weak low-frequency trend is apparent in Sector~60. 

\section{Discussion}

The oscillatory signal detected in the light curve of GJ~3512 differs from the quasi-periodic pulsations (QPPs) that are commonly associated with individual flare events \cite[e.g][]{Balona2015}. Spanning periods of 70--100 minutes, the signal is present in,  {at least, two} separate \emph{TESS} sectors over nearly two years and exhibits amplitudes of 1--2 mmag without evidence of systematic damping. This longevity suggests that the underlying phenomenon is tied to a stable or recurrent magnetic structure on the star rather than to a transient event.

Several alternative interpretations can be discarded. Stellar pulsations in late-M dwarfs have been predicted during pre-main sequence phase, with instability strips located far from the main-sequence locus \citep{RodriguezLopez2019}. With an estimated age of 2--8 Gyr \citep{Morales2019}, GJ 3512 is well beyond the pre-main-sequence phase, making a pulsational origin highly unlikely. Additionally, rotational modulation is inconsistent with the observed timescale; the star's rotation period is approximately 88 days \citep{LopezSantiago2020}, far longer than the 1.5-hour QPP. Similarly, an orbital effect due to the known giant planet GJ~3512b can be excluded because its orbital period is $\sim 203$ days \citep{Morales2025revisiting}, which is orders of magnitude longer than the detected oscillation. Moreover, although the orbit of GJ~3512b is very eccentric ($e = 0.44$), with a semimajor axis $a = 0.34$~a.u., its closest approach to the star is 0.19~a.u. Consequently, a direct magnetic interaction between the star and the planet seems very unlikely \citep{cuntz2000stellar,ip2004star,shkolnik2005hot,zarka2007plasma}.

The most plausible explanation involves magnetohydrodynamic (MHD) oscillations in the stellar corona, analogous to those observed in the Sun and other active stars \citep[e.g.,][]{Nakariakov2009,Kupriyanova2020}. QPPs in solar and stellar flares are commonly attributed to standing slow magnetoacoustic modes in coronal loops or periodic reconnection processes modulated by MHD waves \citep{Wang2021,Lim2022}. The persistence of the modulation in GJ~3512 over years suggests the presence of a long-lived active region capable of hosting stable magnetic loop systems. The observed period drift within the 70--100 minute range may reflect changes in loop length or plasma parameters over time. 

If the apparent correlation between QPP maxima and flare peaks is real, it could indicate that microflaring activity or flare-triggered oscillations contribute to sustaining the modulation. GJ~3512 exhibits a flare frequency of about one event per day, with intense flares every, approximately, two days, consistent with previous studies of mid-to-late M dwarfs \citep[e.g.,][]{Davenport2019,Gunther2020,Stelzer2022}. If the cumulative effect of numerous low-energy flares contributes to maintaining oscillations in large-scale coronal loops, the QPP could represent a signature of quasi-continuous magnetic energy release. Such {recurrent, long-lived} QPPs of this nature have not been reported previously in late-type M dwarfs. Their detection provides an important diagnostic of magnetic topology and energy dissipation in fully convective stars. 

Such sustained pulsation sequence outside of flares indicates the presence of processes, such as reconnection chains \citep{Kumar2025,Schiavo2024} and/or sloshing in long structures, such as loop arcades \citep{Reale2016,Froment2020}, on large scales which keep coherent for extended temporal scales. Arcade loop lengths might not be far from those estimated from analogous \emph{TESS} QPP observations \citep{Ramsay2021}, i.e. around 1000 Mm, but we might be in the presence of a sequence of undistinguishable continuous microflaring activity which triggers the pulsations. It is very interesting that this activity acts in a non-chaotic way, so as to trigger coherent large scale periodic signals.

As already noted \citep{Ramsay2021}, we remark that the waveband of QPPs explained by MHD waves on the Sun \citep{Pugh2016}, i.e., X-rays and EUV, is quite different from the white light of \emph{TESS} in this analysis. QPPs are not observed on the Sun in white light. The reason that reconciles this discrepancy might be simply that the energetics of the \emph{TESS} events is on a much larger scale as compared to the solar one so as to involve deeper and cooler atmospheric layers than the coronal ones.

\section{Conclusions}

We report the discovery of a {recurrent} quasi-periodic signal with a period ranging from 70 to 100 minutes in the optical light curve of the late-M dwarf GJ~3512. This signal was detected across separate \emph{TESS} sectors spanning nearly 2--3 years. This makes it the first case of such a long-lived QPP observed in a late-M dwarf. The signal's persistence and its amplitude of approximately 1--2 mmag suggest an origin in MHD oscillations or periodic magnetic reconnection in long-lived coronal structures, possibly involving one or more stable loops anchored in a persistent active region. Alternative explanations, such as stellar pulsations, rotational modulation, star-planet interaction, or instrumental artifacts, are inconsistent with the observational constraints. This finding provides a new perspective on magnetic activity in fully convective stars by offering insight into coronal loop dynamics and the role of microflaring in energy release. Future multiwavelength observations, particularly in X-rays, are essential to confirm the coronal nature of these oscillations and constrain their physical drivers. The detection of long-lasting QPPs establishes an observational basis for characterizing the timescales, energetics, and coherence of MHD processes in the coronae of fully convective stars, providing constraints that can be directly incorporated into ongoing models of magnetic activity in this stellar regime.

\section*{Acknowledgements}
This work has been funded in part by the Spanish Ministry of Science, Innovation, and Universities and by the State Research Agency (MICIU/AEI/10.13039/501100011033/) under Grants PID2024-158181NB-I00, PID2024-159557OB-C2 and PID2024-159557OB-C21, and by the Regional Ministry of Education, Science, and Universities, Community of Madrid, under Grant IDEA-CM (TEC-2024/COM-89).
J.L-S. acknowledges partial support from the Office of Naval Research (award no. N62909-24-1-2095). FR acknowledges support from Italian Ministero dell'Università e della Ricerca (MUR). GM acknowledges support from ASI/INAF agreement n. 2021-5-HH.1-2022.
The authors thank the reviewer for a useful discussion of the methods and analysis performed in this work. 

\section*{Data Availability}
 
For this work, we used data from the TESS mission. These data can be found in MAST (http://dx.doi.org/10.17909/gsea-a830). 

 



\bibliographystyle{mnras}
\bibliography{GJ3512QPP} 








\bsp	
\label{lastpage}
\end{document}